\documentclass{JINST}
\usepackage{graphicx,amssymb,fontenc,times}
\usepackage{lmodern}

\title{Mapping charge transport effects in thick CCDs with a dithered array of 40,000 stars}

\author{A. Bradshaw$^a$\thanks{Corresponding author.}~, C. Lage$^a$, E. Resseguie$^a$,  J. A. Tyson$^a$\\
\llap{$^a$}University of California, Davis,\\
  Davis, CA  USA  95616\\
E-mail: \email{akbradshaw@ucdavis.edu}}

\abstract{We characterize the astrometric distortion at the edges of thick, fully-depleted CCDs in the lab using a bench-top simulation of LSST observing. By illuminating an array of forty thousand pinholes (30$\mu m$ diameter) at the object plane of a f/1.2 optical reimager, thousands of PSFs can be imaged over a 4Kx4K pixel CCD. Each high purity silicon pixel, 10$\mu m$ square by 100$\mu m$ deep, can then be individually characterized through a series of sub-pixel dithers in the X/Y plane. The unique character [response, position, shape] of each pixel as a function of flux, wavelength, back side bias, etc. can be investigated. We measure the magnitude and onset of astrometric error at the edges of the detector as a test of the experimental setup, using a LSST prototype CCD. We show that this astrometric error at the edge is sourced from non-uniformities in the electric field lines that define pixel boundaries. This edge distortion must be corrected in order to optimize the science output of weak gravitational lensing and large scale structure measurements for the LSST. }


\begin{document}

\section{Introduction}\label{sec:intro}
Precision astrometric performance of the LSST camera will be critically important in order to achieve its scientific goals \cite{lsst-science-book}. The camera will be the largest ever constructed, having a detector surface more than two feet across covered in a mosaic of 189 4Kx4K CCDs totaling more than 3 billion pixels. The individual behavior of each of these pixels, skyscrapers of fully depleted silicon, can significantly differ from the na\"ive square well approximation. Typically, astronomical photometry measurements reduce this fixed-pattern noise by applying dark frame and flat-field corrections to the photometric response. However, some detector non-uniformities affect not only the photometry but also the astrometry of science images. We demonstrate one type of edge non-uniformity near the edges where the potential applied across the CCD changes the way charge is transported from the site of photoconversion to the collecting gate. The square boundary of a pixel's collecting area is distorted, and its position is shifted, leading to the astrometric distortion of point sources.

In this report we use a novel bench-top LSST beam simulator \cite{beamsim-spie} to characterize entire prototype CCDs using an array of 40,000 pinholes of light. Using a series of hundreds of dithered images we probe the performance at the edges of CCD segments. Positional dithering of exposures of the pinhole array is carried out by randomly stepping the CCD orthogonal to the system optical axis by fractional pixels between exposures in a long sequence of exposures.  By comparing pairs of dithers we predict the size of the shift between frames using the entire array of pinholes and compare it to what is measured in some region of interest. In this way we can define the astrometric residual $\vec{\Delta}(x,y)=\vec{x}_{predicted}-\vec{x}_{measured}$ for every location (x,y) on the CCD which has been sampled with a pinhole image during the exposures. The subpixel astrometric distortion near the edges of CCDs and near bloom stops is readily measured with our laboratory setup, and is well described by a simple model. We also find that this astrometric distortion at some of the CCD edges is a function of the illumination level, further complicating the derivation of a correction in scientific data. Besides astrometric distortion at the edges, the detector PSF shape also depends on the exposure level (the so-called brighter-fatter effect \cite{antilogus-brighter-fatter}), wavelength (due to depth of photo-conversion \cite{plazas-bernstein-e-field}), quantum-efficiency variations in the silicon, and the so-called tree rings \cite{decam-treerings} (which are related to pixel boundary shifts). A precise understanding of these effects using empirical measurements will be necessary to fully realize the promise of this colossal camera.

The format of this paper is as follows: in \S~\ref{sec:detection} we briefly describe the lab setup which simulates LSST observing, our exposure strategy, and how we detect the astrometric residual. In \S~\ref{sec:edge model} we discuss the comparison with simple models of the detector edge, and compare to data collected. In \S~\ref{sec:toymodel} we investigate the serial readout edge further, simluating the astrometric error by including a toy model of the pixel boundary shifts in the presence of $\vec{E}$ fields. We conclude in \S~\ref{sec:discussion}, and discuss planned future work.

\section{Detecting the astrometric residual}\label{sec:detection}
\begin{figure}[tbp]
\centering
\includegraphics[width=.8\textwidth]{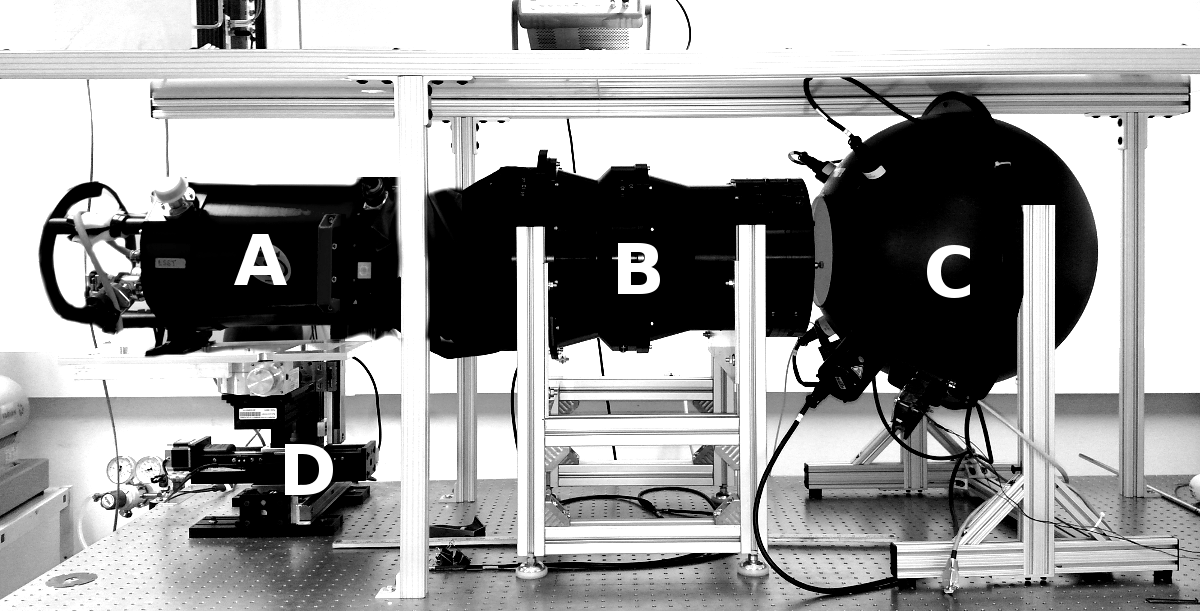}
\caption{The LSST beam simulator assembly. A: camera and cryogenic dewar, B: LSST f/1.2 beam simulator optics and input mask, C: Illuminating sphere, D: precision X/Y/Z stage. }
\label{fig:beamsimpic}
\end{figure}

The LSST beam simulator is an optical bench-top assembly consisting of a camera, beam simulator optics, illuminating sphere, and 3 dimensional stage which is capable of stepping to subpixel accuracy. The automated operation of this equipment enables rapid characterization of CCDs (Figure \ref{fig:beamsimpic}). The optical specifications of the beam simulator were chosen to match that of the LSST, including central obscuration and fast f/1.2 beam. The quality of the telecentric optics is very good, with 50\% energy deposited in a 5$\mu m$ diameter circle, so the optical contribution to the PSF is small when illuminating the camera with a pinhole whose size is on the order of several pixels [the expected atmospheric contribution to the system PSF]. See \cite{beamsim-spie} for more details about the optical performance.

The f/1.2 optical reimaging system is nearly 1:1.  The light from the scattering sphere goes through a band pass filter and the input mask.  These masks are fabricated via a photolithographic process.  Feature accuracy is at the sub-micron level.   For the tests discussed here, we used a mask of 40,000 circular pinholes of 30 micron diameter in a regular grid with 200 micron hole-hole spacing.  We also have masks with sub-pixel size pinholes, which will be used in future exploration of the sub-pixel scale astrometric residuals.

\begin{figure}
\centering
\includegraphics[width=.9\textwidth]{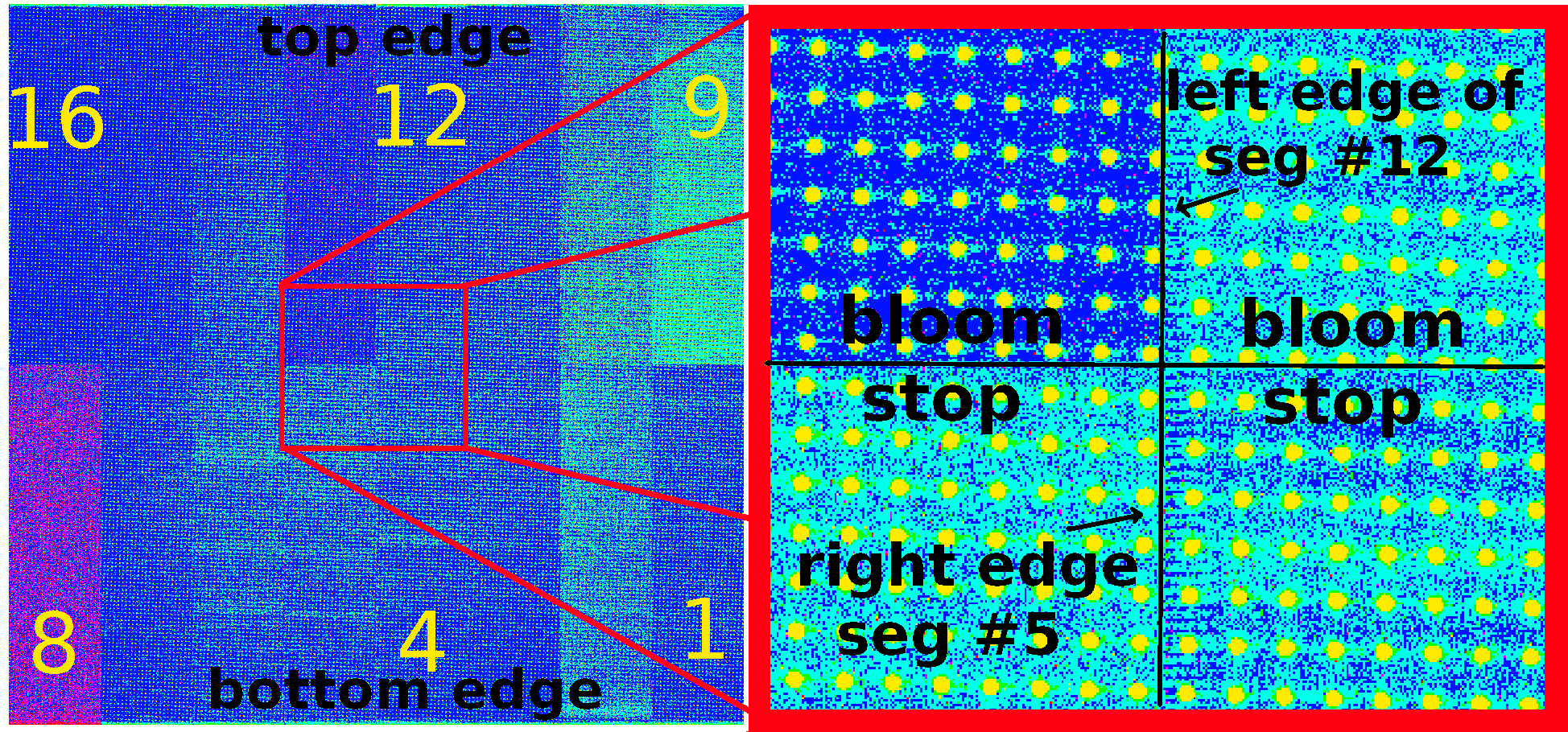}
\caption{Layout of the prototype detector, with bloom stop and edges of segments indicated. The full CCD is illuminated by the 40,000 pinholes in this image. Onthe right, a blowup of a region around the bloom stop and inter-segment edge is shown.  The intentional tilt of the pinhole array is visible. }
\label{fig:ccd-diagram}
\end{figure}

In this report a prototype, early science-grade LSST CCD (manufactured by ITL, model number 1920A) is characterized using our setup. The CCD has 16 independent 1MPix segments in an 8x2 configuration, with each segment having 509 serial columns and 2000 parallel rows as seen in Figure \ref{fig:ccd-diagram}. Across the mid line of the CCD prototype is an anti-blooming drain which is buried in the central channel stop. The illuminating sphere on the far right of Figure \ref{fig:beamsimpic} provides a black body spectrum of white light, which is passed through a changeable bandpass filter to the pinhole array. In this report we use only the R filter, and our pinhole array consists of 30 micron diameter holes spaced 200 microns apart. These 30 micron pinholes produce a nearly Gaussian profile on the CCD with a FWHM of 4 pixels ($\sigma\sim1.5$ pix) when in focus. This width is meant to roughly approximate the pixel sampling of a star in LSST observations.

To investigate the astrometric distortion, we use a precision X/Y/Z stage to first position the surface of the CCD at the focal plane of the LSST beam simulator using a series of exposures in the Z direction (optical axis). Once aligned and focused, the stage and camera are then dithered in the X/Y plane with a series of 300 steps of several pixels, with fractional pixel scatter. The mask is rotated in the XY plane slightly, so that an entire row of pinhiles oversamples a linear feature in the CCD, such as an edge or bloom stop. This dithering strategy is repeated at 6 different illumination levels between 1,000 electrons to 60,000 electrons per peak pixel (less than half full-well).

Bias frame correction was performed using a master bias created from the median of 250 zero second exposures. Each exposure line is median overscan subtracted to remove intermittent line noise (from prototype electronics), after which the master bias frame is subtracted.

\begin{figure} 
\centering
\begin{minipage}{.5\textwidth}
\includegraphics[width=.9\textwidth]{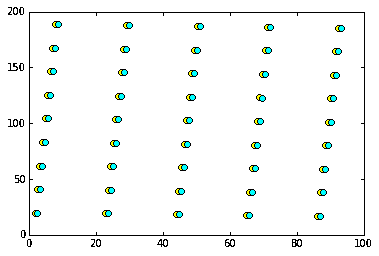}
\end{minipage}\begin{minipage}{.5\textwidth}
\includegraphics[width=.9\textwidth]{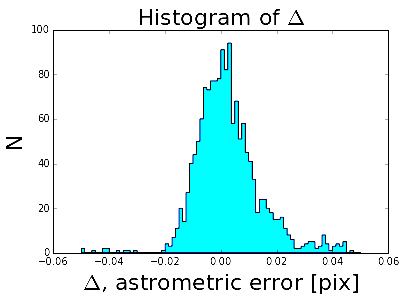}
\end{minipage}
\caption{Imaged pinhole array astrometric distribution. On the left, two consecutive images and the detected pinholes are shown, marked in yellow and blue. On the right, we show a histogram of the centroid shift between such pairs of images. Thousands of pinhole offsets are binned in each CCD segment, excluding the edge pixels, to precisely determine the master shift between two frames. It can be seen that the master dither offset of all pinholes has an error, or width,  of 0.015 pixels. }
\label{fig:shifthist}
\end{figure}

Detection of the pinhole images is then performed using SExtractor \cite{sextractor}, and the resulting multiple catalogs of position, intensity, and shape of each pinhole image in the array are analyzed. For pairs of images, the dither shift between each image is calculated by finding the shift for each individual pinhole and then choosing the median of the distribution of pinholes shifts. This allows a precise shift to be determined, with an error of $\sigma_{shift} \sim .015$ pixels as is shown in Figure \ref{fig:shifthist}. The layout of the 16 segments and a blowup of a region around the bloom stop and inter-segment edge is shown in Figure \ref{fig:ccd-diagram}.  The intentional tilt of the pinhole array can be seen.

\section{Results and modeling the occultation effect}\label{sec:edge model}
\begin{figure} 
\centering
\begin{minipage}{.5\textwidth}
\includegraphics[width=1.\textwidth]{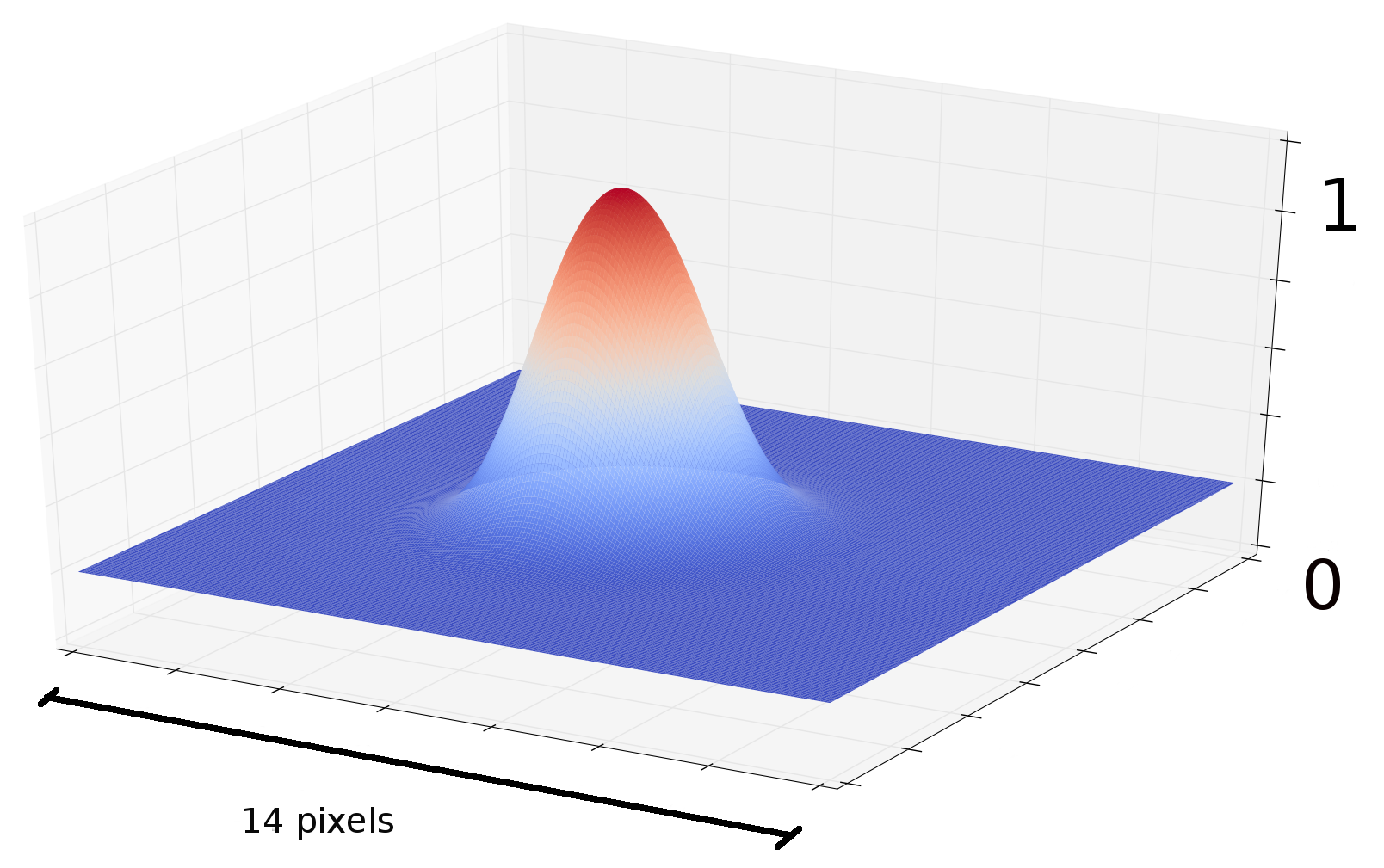}
\end{minipage}\begin{minipage}{.45\textwidth}
\includegraphics[width=1.\textwidth]{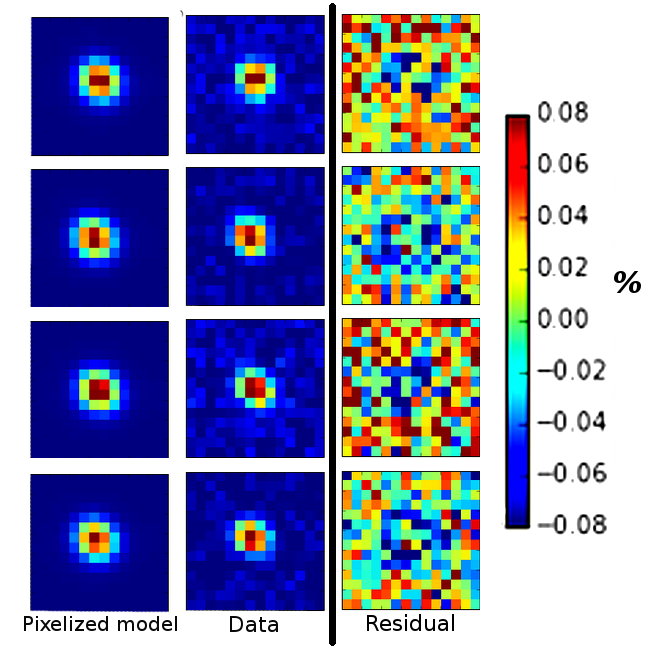}
\end{minipage}
\caption{Forward simulation of the occulation effect. On the left, the drizzled and well sampled PSF image is shown. On the right, a comparison of actual pixelized pinhole images to the model PSF centroided and pixelized at the same location. The far right panels show the difference, which is noisy at the percent level. Each panel is 14x14 pixels. The color bar on the right indicates the residual flux as a fraction of the maximum value (shown for the lowest illumination level $\sim 1k e^-$).}
\label{fig:modelpsf}
\end{figure}

With a known shift between each image pair, it is possible to precisely predict the expected location of a pinhole's image. At the edges of the detector we find the difference between the predicted and measured centroid, the astrometric error $\Delta=x_{predicted}-x_{measured}$, becomes much larger than the positional error $\sigma_{shift}$ for the median of the rest of the frame. Most of this individual pinhole image astrometric error comes from the simple effect of occulting the pinhole image at the edge of the detector, where the centroid is calculated from flux which only partially lays on the detector surface. This creates an abrupt retrograde astrometric error as one approaches any edge of the detector, including at the bloom stop. We modeled this occultation effect using a well sampled PSF from drizzling pinhole images on top of one another (show in Figure \ref{fig:modelpsf}). Our model of the occultaion is made noise free by multiple forward simulations of the data run. By repeated random shifts and pixelizing the drizzled PSF toward the edge we calculate $\Delta=x_{predicted}-x_{measured}$ with a high S/N ratio, using the centroids from SExtractor in the same way as with the real data. 

\begin{figure} 
\centering
\begin{minipage}{.5\textwidth}
\includegraphics[width=.95\textwidth]{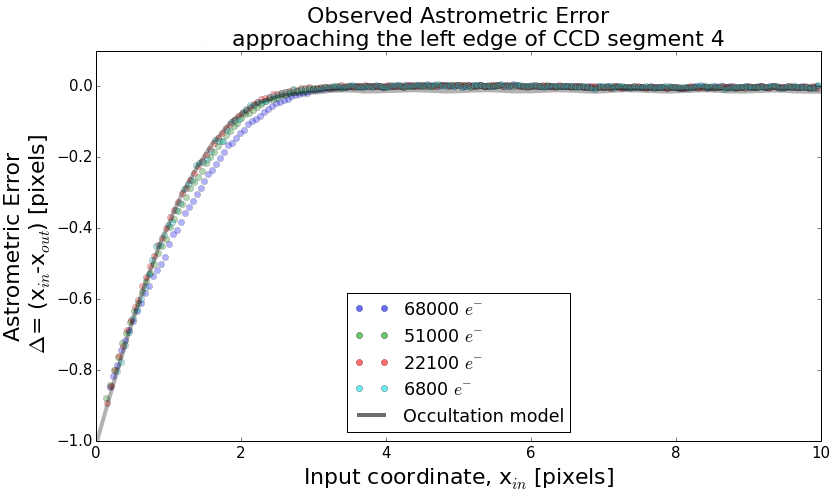}
\end{minipage}\begin{minipage}{.5\textwidth}
\includegraphics[width=.95\textwidth]{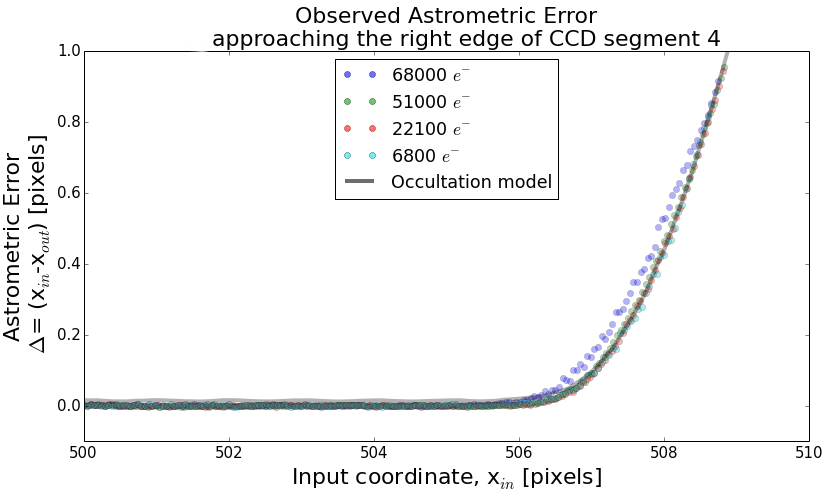}
\end{minipage}
\caption{The astrometric error at the left and right inter-segment edges of segment 4. Astrometric error is dominated by the occultation effect, and the forward simulation model is well matched to the data for all illumination levels below half full well.}
\label{fig:seg4lr}
\end{figure}

\begin{figure} 
\centering
\begin{minipage}{.5\textwidth}
\includegraphics[width=.95\textwidth]{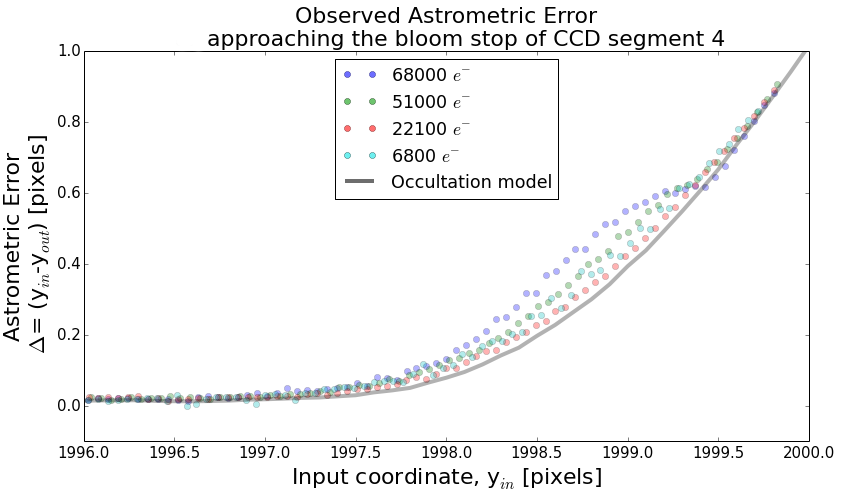}
\end{minipage}\begin{minipage}{.5\textwidth}
\includegraphics[width=.95\textwidth]{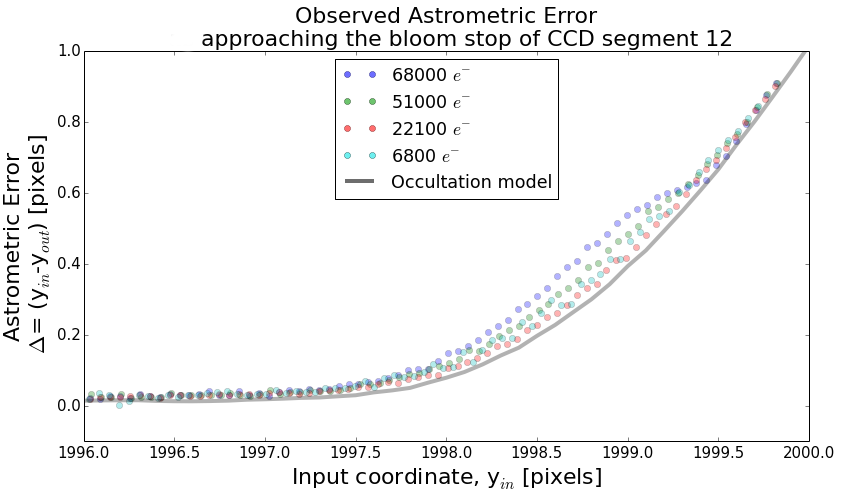}
\end{minipage}
\caption{The astrometric error approaching midline bloom stop from below (segment 4) and above (segment 12). We find a small but significant intensity dependent astrometric residual within several pixels of the bloom stop. }
\label{fig:segbloom}
\end{figure}

\begin{figure} 
\centering
\begin{minipage}{.5\textwidth}
\includegraphics[width=.9\textwidth]{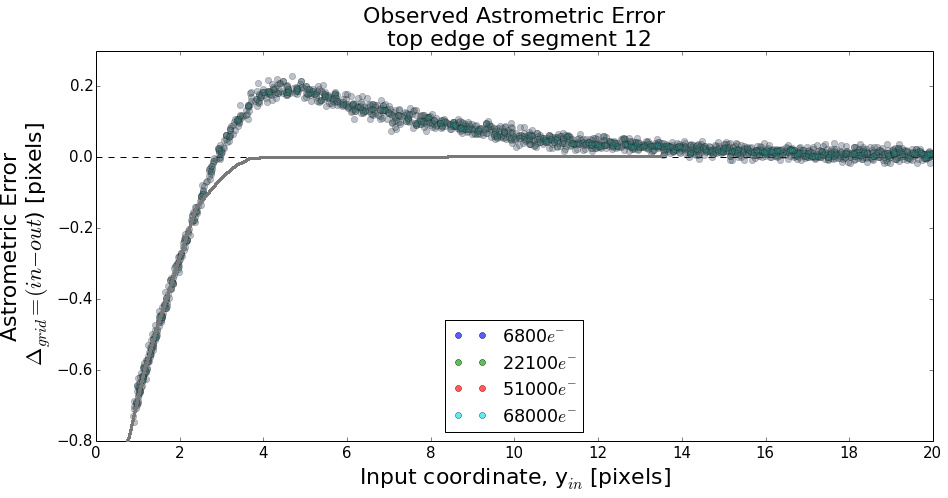}
\end{minipage}\begin{minipage}{.5\textwidth}
\includegraphics[width=.9\textwidth]{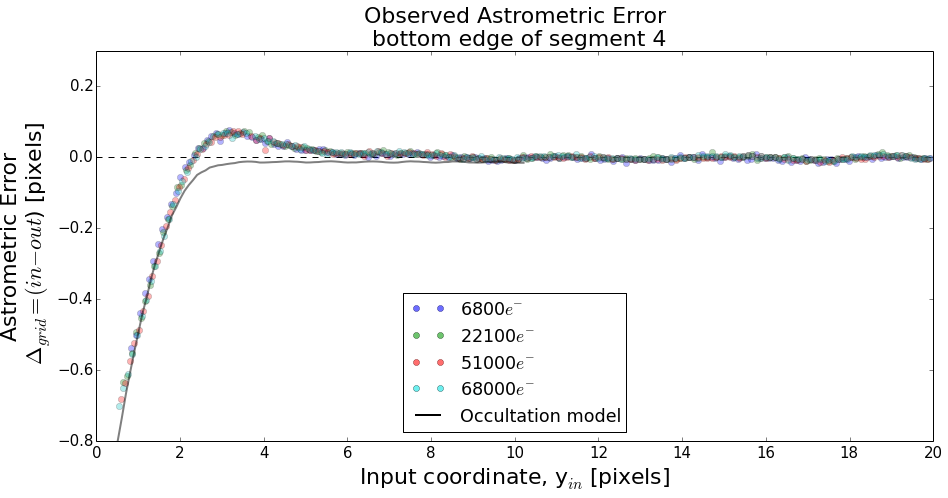}
\end{minipage}
\caption{The astrometric error near the serial readout electronics, located at the top of segment 12 and bottom of segment 4.}
\label{fig:segserial}
\end{figure}

Figures \ref{fig:seg4lr}, \ref{fig:segbloom}, and \ref{fig:segserial} plot the measured astrometric error approaching the edges of segments. The different illumination levels are plotted in different colors. In all these figures, astrometric measurements have been binned (slightly smoothed) in the X direction so that each plotted point is the median of roughly five observed pinholes at that X/Y location. The empirical model of the astrometric error from simple occultation of the PSF, described above, is shown as a gray line in all figures. 

In Figure \ref{fig:seg4lr} it can be seen that at the left and right edges between CCD segments, virtually all of the astrometric error can be attributed to the occultation effect, and the forward simulation model is well matched to the data for all illumination levels. A small divergence from simple occultation is found at higher fluxes, but not as significant as found at the bloom stop and edges of the CCD near the serial registers (discussed below).

Similarly, when approaching the bloom stop in the middle of the CCD (Figure \ref{fig:segbloom}), it can be seen that the astrometric error is again dominated by the occultation effect, but becomes slightly larger than the model at larger illumination levels. Remarkably similar behavior is seen in both segment 4 when approaching the bloom stop from below, and in segment 12 when approaching it from above. Such behavior may be expected due to the bloom stop effectively draining electrons and thus driving the observed PSF closer to the midline.

When approaching the serial registers (Figure \ref{fig:segserial}) the astrometric error differs significantly from the simple occultation model. The astrometric error is first positively increasing as pixel 0 is approached, indicating that the measured PSF is prograde [i.e. pinhole image appears pulled toward the serial register] to the predicted location, before then being occulted at the edge. This positive astrometric error, in the case of segment 12, has an amplitude of 0.1 pixels at 10 pixels away from the edge. We hypothesize that this prograde shift is due to a non-uniform field lines across the thickness of the detector distorting the nominal boundaries for a pixel. The non-uniform field lines can be sourced by positive voltages applied at the scupper and serial register at the top and bottom of the CCD. Ordinarily a photoelectron travels downward through the 100 microns of silicon, only diverting in the last few microns toward the pixel's collecting region. In the presence of outside $\vec{E}$ field lines this downward charge transport may become slightly transverse, pulling electrons toward outer electronics. The effective boundaries of a pixel near these readout electronics is thus broadened slightly, so that light which is incident away from the edge may have its flux measured as closer to the edge (corresponding to a positive $\Delta$ near these edges, as measured). Curiously, the amplitude and onset of this astrometric distortion for this test device is different at the top and bottom of the CCD, an effect for which we at present have no explanation.

\section{Simulation of charge transport}\label{sec:toymodel}

\begin{figure} 
\centering
\begin{minipage}{.5\textwidth}
\includegraphics[width=.9\textwidth]{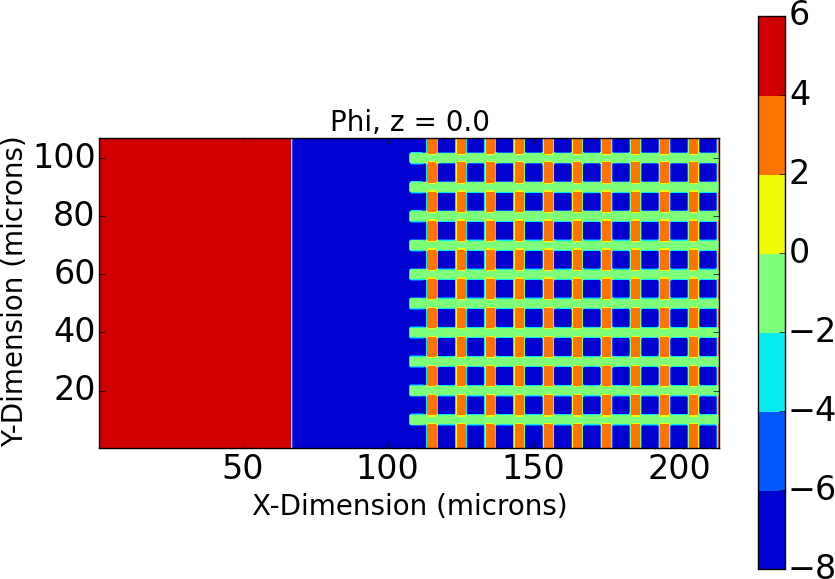}
\end{minipage}\begin{minipage}{.5\textwidth}
\includegraphics[width=.9\textwidth]{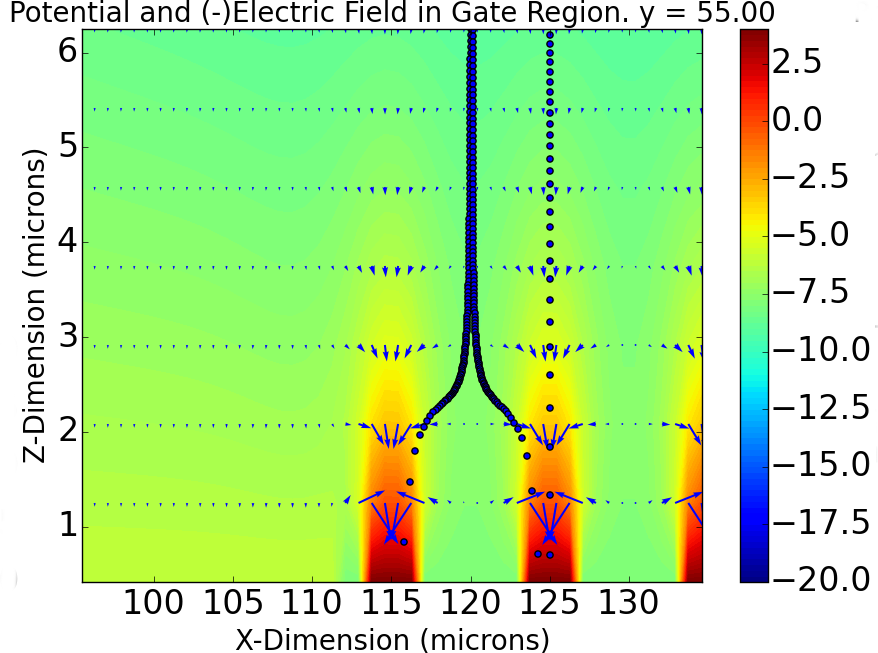}
\end{minipage}
\caption{Toy model of a photo electron drift in a CCD. On the left the potential is mapped, with color code indicating the amplitude of the realistic voltages applied. On the right a vertical cross-section of the drift of electrons. Field lines indicate direction of drift, while the color again indicates potential. The vertical lines show how electrons placed at the boundary of a pixel drift toward their respective collecting regions.}
\label{fig:poisson-model}
\end{figure}

We are developing a 3D model of charge transport, using known CCD polysilicon geometry and known applied voltages. For this paper, using a Poisson toy model of the electron drifting toward collection, we simulate the effect of nearby realistic voltages on the electron path.  To perform this simulation, we solve Poisson's equation using multi-grid methods on a region 128 microns x 256 microns in area and 100 microns thick.  The unit cell is 0.5 microns on a side.  A uniform potential of -50V is applied on the back (light-collecting) side, and the appropriate potentials (as seen in the left side of Figure \ref{fig:poisson-model}) are applied to the front side.  The channel-stop regions are represented by charges corresponding to the channel-stop implants.  Using this simulation we are able to map a given input location for photoelectrons to the pixel where it will be measured. Figure \ref{fig:poisson-model} illustrates the simulation of pixel drift.

\begin{figure} 
\centering
\includegraphics[width=.64\textwidth]{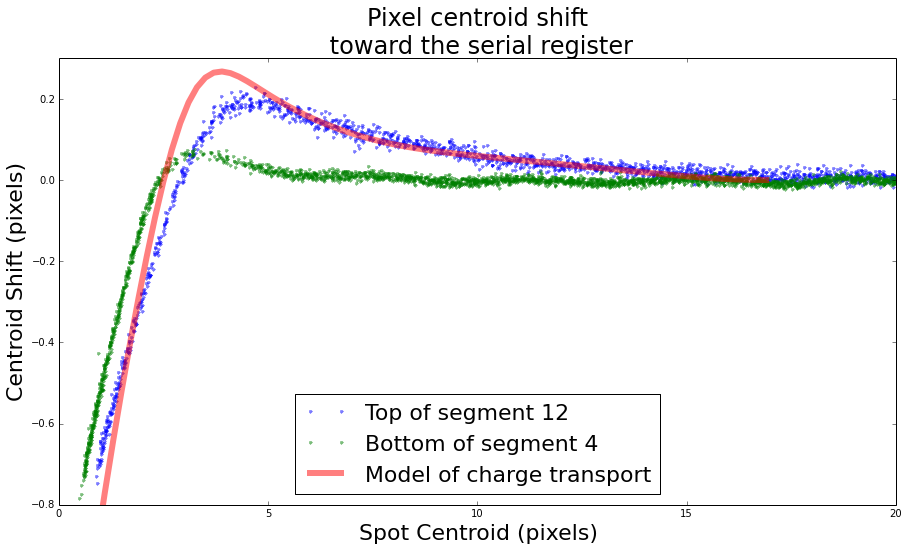}
\caption{The measured astrometric error at the top and bottom of the CCD near the serial registers, plotted as points. The data are in qualitative agreement with the edge model including pixel boundary distortions. The toy model of charge transport is not a fit to either of the data sets.}
\label{fig:moneyplot}
\end{figure}

Using the distorted pixel boundaries we can once again shift-and-pixelize a model PSF toward the CCD edge as done before, but using the distorted pixel boundaries from the toy model instead of regularly spaced ones. This yields an estimate of the astrometric error due to both occultation and distorted pixel boundaries. The shape and amplitude of this modeled astrometric error is shown in Figure \ref{fig:moneyplot}, and agrees well with the amplitude and onset of the astrometric distortion measured at the top and bottom of the CCD.  Note that the model has not been fit to the data in this case, but instead is meant to show the qualitative agreement between the model and measured data.  We plan to make this model more quantitative in the future, and will use the improved model together with more extensive measurements to help us understand and potentially minimize the astrometric error.

\section{Discussion}\label{sec:discussion}
Measurements made using the LSST beam simulator laboratory are beginning to shine light on the low level systematics lurking inside of our CCDs. Using a series of dithered images of a pinhole array, we show that the edges of segments of the CCD behave in irregular but well defined ways which may systematically bias cosmological measurements. The astrometric errors due to non-uniform field lines are detected at high signal to noise, allowing precise comparison with various physically motivated models and opening the possibility of correcting scientific data using these models in the future.  

There are likely many more low-level systematics waiting to be unveiled, quantified, and corrected. Further characterization of prototype CCDs will investigate the photometric and astrometric performance of the devices as a function of backside bias, filter (wavelength), and sky level. In addition, variously designed pinhole arrays will enable studies of individual pixel response, informing our model.  In the end, we hope to arrive at a sufficient understanding of the astrometric and photometric effects on all spatial scales in these devices, to inform pipeline software correction to first order. Moreover, using these corrections we look forward to testing shape-fitting routines on realistic data before first light on the LSST.

\acknowledgments
We thank Perry Gee, Kirk Gilmore, and John Warren for their help. We acknowledge partial support from DOE grant DE-SC0009999.

\end{document}